\documentstyle[preprint,aps]{revtex}

\begin{document}

\draft

\preprint{$
\begin{array}{l}
\mbox{UCD--96--18}\\[-3mm]
\mbox{AMES-HET-96-03}\\[-3mm]
\mbox{May 1996}\\
\end{array}
$}

\title{Searching for $t \rightarrow c g$ at the Fermilab Tevatron}

\author{T. Han$^a$, K. Whisnant$^b$, B.-L. Young$^b$ and X. Zhang$^{b,c}$ }

\address{{ $^a$Department of Physics, University of California, Davis,
 CA 95616, USA}\\
{$^b$Department of Physics and Astronomy,
 Iowa State University, Ames, IA 50011, USA}\\
{$^c$Institute of High Energy Physics, Academia Sinica,
Beijing, P.R. China}\\
}
\maketitle

\begin{abstract}

We examine the experimental observability of the decay mode $t
\rightarrow c g$ at the Fermilab Tevatron via the flavor-changing
neutral current vertex $t \bar c g$. We find that with
the existing data, one should
be able to probe the $t \bar c g$ coupling to a value
smaller than indirect limits previously obtained from
$b\rightarrow s \gamma$ and the measured branching fraction for
$t\rightarrow bW$, reaching $BF(t \rightarrow cg) \sim$ 15\% - 28\%.
A data sample of 1~fb$^{-1}$ (10~fb$^{-1}$) at $\sqrt{s}=2$~TeV may
probe the $t \rightarrow c g$ branching fraction to a level of 2\%
(0.5\%).

\end{abstract}
              
\narrowtext

\vfill
\eject

\section{Introduction}

With the discovery of the top quark \cite{cdfd0} the long anticipated
completion of the fermion sector of the standard model has been
achieved. Its unexpected large mass in comparison with the other known
fermions suggests that the top quark will play a unique role in probing
new physics, and has prompted both theorists and experimentalists alike
to search for anomalous couplings involving the top quark. On the
experimental side, the CDF \cite{incandela,lecompte} and D0 \cite{APH}
collaborations have begun to explore the
physics of top quark rare decays, and
interesting bounds on flavor-changing decays to electroweak gauge bosons
have been reported \cite{lecompte}. On the theoretical side, a
systematic examination of anomalous top quark interactions
is being actively undertaken \cite{hwyz,treview}.

An interesting set of anomalous interactions are those given by the
flavor-changing chromo- and electro-magnetic operators
\begin{equation}
 \frac{\kappa_g}{\Lambda} g_s{\overline c}\sigma^{\mu\nu}
 \frac{\lambda^a}{2} t G^a_{\mu\nu} + h.c.
 \label{eq:tcglue}
\end{equation}
and
\begin{equation}
 \frac{\kappa_\gamma}{\Lambda} e {\overline c} \sigma^{\mu\nu} t
 F_{\mu\nu} + h.c.,
 \label{eq:tcgamma}
\end{equation}
where $\Lambda$ is the new physics cutoff, $\kappa_g$ and $\kappa_\gamma$
define the strengths of the couplings, and $G_{\mu\nu}$ and $F_{\mu\nu}$
are the gauge field tensors. The investigation of these couplings is
well motivated. Although these operators can be induced in the standard
model by high order loops, their effects are too small to be observable
\cite{tcgammaSM}. However, it has been argued that they may be enhanced
significantly in many extensions of the standard model, such as SUSY
or other models with multiple Higgs doublets \cite{tcgammaSM,multi},
models with new dynamical interactions of the top quark \cite{dyn}, and
models where the top quark has a composite \cite{comp} or soliton
\cite{sol} structure. Therefore, any observed signal indicating these
types of couplings is direct evidence for non-standard physics and
will improve our understanding of flavor dynamics.

In this letter we propose an optimized procedure to search for the
coupling $t \bar c g$ via the decay $t \rightarrow cg$ at the Tevatron
energies. We find that an improved limit on $\kappa_g$, better than that
obtained before from indirect constraints \cite{hwyz}, is possible
even based on the existing 200~pb$^{-1}$ Tevatron data in the $W+3$-jet
mode of $t$-$\bar t$ pair production.

\section{Existing Limits on Flavor-Changing Top Quark Couplings}

The strategy of searching for anomalous top quark couplings consists of
two complementary approaches: (i) obtain indirect bounds from low energy
processes in which top quark anomalous couplings can enter via
loop processes, or from the bound on $t \rightarrow bW$ which
puts limits on other decay modes of the top, and (ii) direct searches
at high energies for the effects of the anomalous couplings in top quark
production and decay. We have found earlier \cite{hwyz} that the
experimental lower limit on $BF(t \rightarrow bW)$ from CDF
\cite{incandela} implies an upper bound of $BF(t \rightarrow cg)<0.45$
at one standard deviation, which gives the limit $|\kappa_g| < 0.95$
for $\Lambda$~=~1~TeV. Then from data on $b \rightarrow s + \gamma$
\cite{cleo}, we found correlated bounds on $\kappa_g$ and $\kappa_\gamma$,
with $|\kappa_\gamma| < 0.3$ for $|\kappa_g|<0.95$ and
$\Lambda$~=~1~TeV. More recently the CDF data \cite{lecompte} gives the
bound
\begin{equation}  
  BF(t \rightarrow c \gamma) + BF(t \rightarrow u \gamma) < 2.9\%  
  \label{eq:tcgammadata}
\end{equation}
at 95\% Confidence Level (CL), which translates to $\kappa_\gamma/\Lambda < 
0.73/\sqrt{BF(t \rightarrow b W)}$, where $\Lambda$ is in units of TeV.  
Reference \cite{lecompte} also gives the bound
\begin{equation}
  BF(t \rightarrow c Z) + BF(t \rightarrow u Z) < 90\%  
  \label{eq:tcZdata}
\end{equation}
at 90\% CL, which puts limits on the flavor-changing neutral current
couplings $tcZ^0$ and $tuZ^0$.  Using the anomalous coupling
$\kappa_{tc}$ defined in Ref. \cite{hpz}, which denotes the combined
effect of $V+A$ and $V-A$ $Zt\bar c$ couplings, we obtain a rather
loose limit of $\kappa_{tc} < 1.3/\sqrt{BF(t \rightarrow b W)}$. The
low energy data give $\kappa_{tc} < 0.29$ \cite{hpz}.
Hence for the anomalous couplings $\kappa_\gamma$ and $\kappa_{tc}$ the
low energy data still provide tighter bounds than the direct limits from
top quark decay. 

The situation for the anomalous coupling $\kappa_g$ is different.
Because the indirect constraint $|\kappa_g|<0.95$ is relatively
weak, and since the interaction in Eq.~(\ref{eq:tcglue}) involves the strong
coupling constant, there is the possibility that it can contribute
significantly to top quark production and decay in hadron colliders.

\section{Top quark decay to charm-gluon at the Tevatron}

At the Fermilab Tevatron, the cross section for $t\bar t$ production is
about 5~pb at $\sqrt{s}=1.8$~TeV. The CDF and D0 experiments have each
collected about 100~pb$^{-1}$ of data. Since a
significant branching fraction for $t \rightarrow cg$ is still allowed,
the current data sample should be sufficient to put improved limits on
$\kappa_g$. Because QCD backgrounds are very large at hadron colliders,
it is best to look for events where one top quark decays
semi-leptonically $t \rightarrow W^+ b \rightarrow \ell^+ \nu b$ (or
$\bar t \rightarrow W^- \bar b \rightarrow \ell^- \bar\nu \bar b$)
($\ell = e$ or $\mu$) and the other decays $\bar t \rightarrow \bar c g$
(or $t \rightarrow c g$), i.e.,
\begin{eqnarray}
p \bar p \rightarrow t \bar t \rightarrow
\ell^+ \nu b \bar c g \ {\rm ~or~} \ \ell^- \bar\nu \bar b c g.
\label{Signl}
\end{eqnarray}
The signature is then $W(\rightarrow \ell^\pm \nu)+3$~jets, where
two jets ($cg$) reconstruct to the top mass, and  so do
the other jet  ($b$) and the $W$.

To obtain the signal event rate, we calculate the top-quark pair
production via $q \bar q \rightarrow t \bar t$ and $gg \rightarrow t
\bar t$ using the lowest order matrix elements, and normalize the total
cross section to theoretical results which include order $\alpha_s^3$
corrections \cite{as3} by including a $K$-factor of 1.4.  For the parton
distributions we use the recent parametrization MRS Set-A
\cite{mrsa}. The top decays are calculated using exact matrix elements
for each decay, assuming an on-shell $W$. We have ignored the top quark
spin correlations since the top-quark production mechanisms we consider
give insignificant top-quark polarization \cite{nopol}. The top-quark
branching fractions may be obtained from the partial width ratio \cite{hwyz}
\begin{equation}
{\Gamma(t \rightarrow cg) \over \Gamma(t \rightarrow bW)} =
{64 \sqrt2 \pi \alpha_s \over 3 G_F \left(1-{M_W^2\over m_t^2}\right)^2
\left(1+2{M_W^2\over m_t^2}\right)} \left({\kappa_g\over \Lambda}\right)^2 =
0.91 \kappa_g^2,
\label{eq:BRtcg}
\end{equation}
and thus, assuming that only $t \rightarrow bW$ and $t \rightarrow cg$
contribute dominantly to top decays,
\begin{equation}
BR(t \rightarrow cg) =  {0.91\kappa_g^2 \over 1 + 0.91\kappa_g^2 },
\label{eq:BR}
\end{equation}
where we have taken $m_t = 175$~GeV, $\alpha_s(2m_t)=0.099$,
$\Lambda$~=~1~TeV, and the masses of the $b$ and $c$ quarks have been
ignored. The signal cross section for the reaction given in
Eq.~(\ref{Signl}) can then be written as
\begin{equation}
\sigma = \sigma_{max}
{(1.82)^2(\kappa_g/0.95)^2 \over [1 + 0.82(\kappa_g/0.95)^2]^2},
\label{eq:sigma}
\end{equation}
where $\sigma_{max}$ is the signal cross section for the maximum allowed
value of the $\kappa_g$ (=0.95). 

Without requiring $b$-tagging to begin with, we can identify the jet
which goes with the $W$ as follows: the pair of jets which best
reconstructs the top mass is assumed to come from the $t \rightarrow cg$
decay, and the other jet is then identified as the $b$-quark
jet. Although the transverse momentum of the neutrino can be taken as
the missing $p^{}_T$, there is a two-fold ambiguity in determining the
neutrino momentum along the beam direction \cite{pnu}. We choose the
solution which best reconstructs the top mass using the momenta of the
jet previously identified as the $b$ jet, the charged lepton, and the
neutrino.

To make the calculation more realistic, we 
simulate the detector effects by assuming a Gaussian energy smearing
for the electromagnetic and hadronic calorimetry as follows:
\begin{eqnarray}
\Delta E/E &=& 30\%/\sqrt E \oplus 1\%, \quad  
{\rm for \ \ leptons} \nonumber\\
&=&  80\%/\sqrt E \oplus 5\%, \quad  {\rm for \ \  jets,} 
\label{smear}
\end{eqnarray}
where the $\oplus$ indicates that the $E$-dependent and $E$-independent
errors are to be added in quadrature, and $E$ is to be measured in GeV.

The dominant background is from $W$ production plus three 
QCD jets \cite{wjjjcode},
\begin{eqnarray}
p \bar p \rightarrow  W^\pm  jjj\rightarrow \ell^\pm \nu jjj .
\label{Bckgnd}
\end{eqnarray}
Although the production rate of the background process is 
significantly larger
than that of $t\bar t$ production at Tevatron energies, 
the kinematics for those processes
is quite different, especially after imposing the top-quark 
mass constraint.

To simulate the detector coverage and help reduce the background, 
we first impose the following ``basic''
acceptance cuts on the transverse momentum ($p^{}_T$), pseudo-rapidity
($\eta$), and the separation in the azimuthal angle-rapidity plane
($\Delta R$) of the charged lepton, jets and missing transverse
momentum
\begin{equation}
 p_T^\ell, p_T^{miss} > 15~{\rm GeV}, \ \ p_T^j> 20~{\rm GeV}, \qquad
 |\eta^\ell|, |\eta^j|<2.5, \qquad 
 \Delta R_{\ell j} > 0.4, \ \ \Delta R_{jj} > 0.8.  
\label{EQ:BASIC}
\end{equation}
The higher transverse momentum and $\Delta R$ cuts for the jets is
motivated by the hard nature of the heavy top decay. 
The signal cross section for the maximal allowed value of
$\kappa_g$ is reduced from about 550~fb with no cuts to around 322~fb
with these basic cuts, while the $W+3$~jets background is about 10.7~pb
after the cuts.

To improve the relative strength of the signal, we make use of
the following facts:
\begin{itemize}
\item
the top-antitop invariant mass $M(t \bar t)$ has a kinematical lower limit
($2m_t$ before energy smearing is applied), while the lower limit is
significantly smaller for the background, near the $Wjjj$
threshold;
\item  
the final state jets in the signal have transverse momenta typically the 
order of $\frac{1}{2} m_t \simeq 80$ GeV due
to the nature of top-quark two-body decay,
while all the jets in the background events
tend to be soft. We can define two scalar sums of the 
transverse momenta:
\begin{equation}
p^{}_T(j_1j_2) \equiv |\vec p^{j_1}_T|   + |\vec p^{j_2}_T|,  \qquad
p^{}_T(jjj) \equiv p^{}_T(j_1j_2) + |\vec p^{j_3}_T|,
\label{EQ:PTscalar}
\end{equation}
where the jet transverse momenta are ordered such that $|\vec p^{j_1}_T|
> |\vec p^{j_2}_T| > |\vec p^{j_3}_T|$. In fact, the signal spectra are much
harder than the background at the low end, but they are limited by the
physical scale $2m_t$, so that the background tends to extend relatively
further at the high end. 
\end{itemize}
With these points in mind, we therefore accept events with
\begin{equation}
M(t\bar t) > 2m_t = 350 \ {\rm GeV}, \qquad
100 < p^{}_T(j_1j_2) <  300 {\rm~GeV}, \qquad
150 < p^{}_T(jjj) < 400 {\rm~GeV}.
\label{EQ:LEVELII}
\end{equation}
If we define our Level-II cuts as those given in Eq.~(\ref{EQ:LEVELII}),
then the maximal signal is reduced only moderately to
about 275~fb, while the background is reduced by about a factor of six
to around 1870~fb. 

In Fig.~1 we show the reconstructed top-quark mass distributions
$M(cg)$ and $M(bW)$ after making both the basic and Level-II cuts,
where the $W$ momentum is obtained from the momenta
of the charged lepton and the reconstructed neutrino.
We see from Fig.~1(a) that the continuum background is still above
the $M(cg)$ signal peak in the region around $m_t$. A further
improvement can be made if we impose the cut
\begin{equation}
|M(cg)-m_t| < 20{\rm~GeV}.
\label{EQ:MCG}
\end{equation}
In Fig.~1(b) the dashed histograms show how the background in the
$M(bW)$ distribution is reduced by the cut in Eq.~(\ref{EQ:MCG}). The
signal distribution is reduced only moderately by this cut (see the
dotted and the solid curves) and is now
within a factor of two of the background when $\kappa_g$ is at its
maximal value. The signal observability can be maximized if we consider
the events in the mass range
\begin{equation}
|M(bW)-m_t| < 30 {\rm~GeV}.
\label{EQ:MBW}
\end{equation}
After the final cut in Eq.~(\ref{EQ:MBW}), the maximal signal cross
section is
\begin{equation}
\sigma_{max} = 195 {\rm~fb}
\label{eq:sigmamax}
\end{equation}
and the background is about 400~fb. Therefore up to 40 signal events
would be expected for the current integrated luminosity, with a
background of about 80 events, which would correspond to nearly a $4\sigma$
signal near the $M(bW)$ peak. Table~\ref{T:ONE} summarizes the effect of
the various cuts on the maximal signal and the background. The signal
rate for non-maximal $\kappa_g$ is easily computed using
Eqs.~(\ref{eq:sigma}) and (\ref{eq:sigmamax}).

We have so far optimized $S/B$ only based on kinematical variables.
If we further require a $b$-tagging on the jet that satisfies
Eq.~(\ref{EQ:MBW}) in top-quark mass reconstruction, 
and assume a 50\% $b$-tagging
efficiency and 1\% impurity \cite{btag}, one expects to improve
the $S/B$ ratio by a factor of 50.
 
To estimate the sensitivity to $\kappa_g$ for a given integrated
luminosity, we can take the cross section given in Eq.~(\ref{eq:sigma}),
using the value of $\sigma_{max}$ in Eq.~(\ref{eq:sigmamax}), and compare
it to the background rate. With Gaussian statistics, a measurement
is sensitive to the signal at 99\%~CL when
\begin{equation}
S/\sqrt{S+B} = 3.
\label{EQ:GAUSS}
\end{equation}
The solid line in Fig.~2 presents the anomalous coupling $\kappa_g$
versus the integrated luminosity required at the Tevatron with
$\sqrt{s}=1.8$~TeV. The dashed curve in Fig.~2 shows the improvement in
sensitivity if $b$-tagging is employed.  We see that with 200 pb$^{-1}$
integrated luminosity as accumulated by CDF and D0 collaborations, one
should be able to probe this anomalous coupling to $\kappa_g \sim 0.43 - 0.65$
with or without $b$-tagging, corresponding to a branching fraction
$BR(t \rightarrow cg) \sim$ 15\% - 28\%.  In other words, if we assume
the anomalous coupling $\kappa_g$ is naturally of order unity and allow
the new physics cutoff scale ($\Lambda$) to change, then the current
Tevatron data should be sensitive to $\Lambda \sim 2$ TeV.

In the future, with $\sqrt{s}=2$~TeV and
the expected 1~fb$^{-1}$/yr integrated luminosity of
the Main Injector, or 10~fb$^{-1}$/yr at the Tevatron Upgrade, 
further dramatic improvements in the
limits on $\kappa_g$ should be possible. Potential results for
the 2 TeV Tevatron are also shown in Table~\ref{T:ONE} (in parentheses)
and Fig.~2 (dotted and dash-dotted curves).
We see that a branching fraction of order 2\% (0.5\%) would be
reached for 1~fb$^{-1}$ (10~fb$^{-1}$) integrated luminosity,
corresponding to a probe of the coupling down to $\kappa_g = 0.15 (0.07)$.

\section{Discussion and Summary}

Before concluding, a few remarks are in order. First, the top-quark
events with the SM hadronic decay $t \rightarrow bW \rightarrow bjj$
may pose a background to our signal as well if one of the jets escapes
detection. However, our requirement in Eq.~(\ref{EQ:MCG}) for
$m_t$ reconstruction by two non-$b$ jets would hopefully remove
this background. Second, if such a coupling exists at an observable level, 
there might also be a possibility of significant 
single top production via the anomalous vertex.
A study of $q\bar q \rightarrow t\bar c$ has recently appeared \cite{tcg}.
Given the fact that the signal from the decay $t \rightarrow cg$ is more
kinematically characteristic, our results here should be more 
promising. More detailed studies with all contributing processes which
involve the $t \bar c g$ coupling, including calculations at LHC
energies, will be presented elsewhere \cite{hhwyz}.

In summary, we have found that the current Tevatron data sample can
already be used to improve the current limit on (or detect the existence
of) an anomalous flavor-changing magnetic $t\bar c g$ coupling via a
direct search for $t \rightarrow cg$ decay in top-antitop pair production. 
The upgraded Tevatron will allow a probe of the $t \rightarrow cg$
branching fraction to the order of 1\%.

\section{Acknowledgments}

This work was supported in part by  the U.S.~Department 
of Energy under Contracts  DE-FG03-91ER40674 (T. Han) and DE-FG02-94ER40817 
(K.Whisnant, B.-L. Young and X.Zhang).  
  
\vfill
\eject

\vfill
\eject


\begin{table}[htb]
\centering
\caption[]{Cross sections in units of fb for the 
$t\bar t \rightarrow \ell^\pm \nu bcg$ signal 
(with maximal coupling $\kappa_g=0.95$),
and the $Wjjj$ background at the Tevatron with $\sqrt{s}=1.8$~TeV
(2~TeV). The results are shown at various stages of the analysis: for the
basic acceptance cuts of Eq.~(\ref{EQ:BASIC}), after the Level-II cuts of
Eq.~(\ref{EQ:LEVELII}), after the cuts on $M(cg)$ and
$M(bW)$, and finally after including $b$-tagging. A 50\% $b$-tagging
efficiency and 1\% impurity are assumed \cite{btag}. }
\begin{tabular}{|c|c|c|}
Cuts & Signal: $\sigma_{max}(t\bar t \rightarrow \ell^\pm \nu bcg)$ (fb) & 
Background:  $\sigma( W^\pm jjj\rightarrow \ell^\pm \nu jjj)$ (fb)  \\ \hline
Basic Cuts& 322  (447) &  7920 (10700) \\ \hline
Level-II & 276 (380) & 1870 (2850) \\ \hline
$|M(cg)-m_t|<20$~GeV & 239 (329) & 767 (1150) \\ \hline
$|M(bW)-m_t|<30$~GeV & 195 (268) & 399 (585) \\ \hline
plus $b$-tagging & 98 (134) & 4 (6) \\
\end{tabular}
\label{T:ONE}
\end{table}


\bigskip
\bigskip
\centerline{FIGURE CAPTIONS}

FIG.~1 Invariant mass distributions after basic and Level-II cuts at the
Tevatron with $\sqrt{s}=1.8$~TeV for (a) $M(cg)$ and (b) $M(bW)$ with
the additional cut $|M(cg)-m_t|<20$~GeV. The solid curves are the signal
and the dashed curves represent the $W+3$~jet background. In the $M(bW)$
distribution, also shown are the background (upper dashed) and signal
(dotted) curves before the $M(cg)$ cut (the
signal is about 20\%  lower with this cut).

\bigskip
FIG.~2 Sensitivity to $\kappa_g$ vs. integrated luminosity at the
Tevatron at 99\% CL. The solid (dashed) curves represent the sensitivity
at $\sqrt{s}=1.8$~TeV without (with) $b$-tagging, and the dotted
(dot-dashed) curves represent the sensitivity at $\sqrt{s}=2$~TeV
without (with) $b$-tagging.

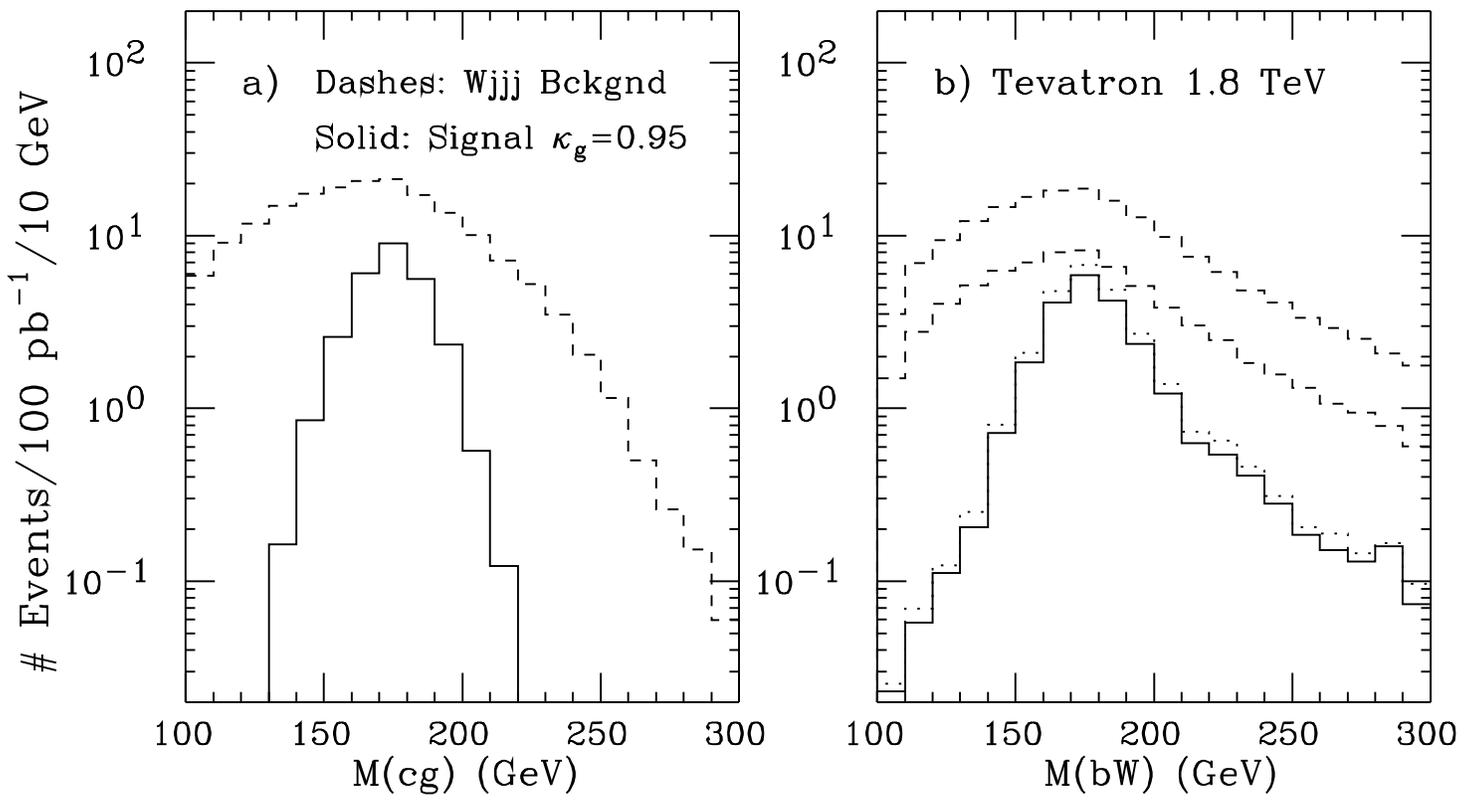
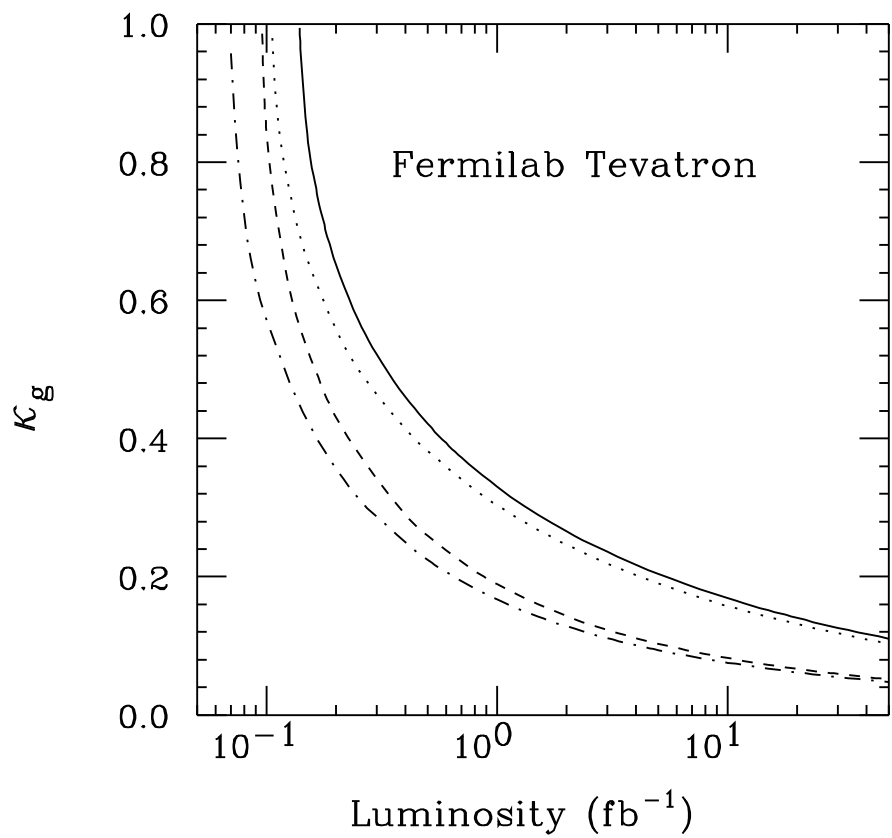
\end{document}